\documentclass[floats,aps,showpacs,twocolumn]{revtex4}

\usepackage[dvips]{graphicx}
\usepackage{amsfonts}

\newcommand{\tmin}{\tau_{\text{min}}}
\newcommand{\xmin}{x_{\text{min}}}

\newcommand{\geqa}{\stackrel{>}{\scriptstyle \sim}}

\newcommand{\ef}{\epsilon_{\text {\tiny F}}}
\newcommand{\ecb}{\overline{\cal E}}
\newcommand{\ect}{\widetilde{\cal E}}
\newcommand{\ob}{\overline{\Omega}}
\newcommand{\ot}{\widetilde{\Omega}}
\newcommand{\rmb}{\rho_{\text {\tiny MB}}}
\newcommand{\rhob}{\overline{\rho}}
\newcommand{\rhot}{\widetilde{\rho}}
\newcommand{\ec}{{\cal E}}
\newcommand{\sca}{{\cal S}}
\newcommand{\scab}{\overline{{\cal S}}}
\newcommand{\scat}{\widetilde{{\cal S}}}
\newcommand{\dc}{{\cal D}}
\newcommand{\nc}{{\cal N}}
\newcommand{\ncb}{\overline{\cal N}}

\newcommand{\mub}{\overline{\mu}}
\newcommand{\mut}{\widetilde{\mu}}
\newcommand{\tb}{\overline{T}}
\newcommand{\tti}{\widetilde{T}}
\newcommand{\taut}{\tau_{\scriptscriptstyle T}}
\newcommand{\xh}{x_{\text{H}}}

\newlength{\figwidth}
\begin{document}

\title{Fluctuations in the level density of a Fermi gas}

\author{P. Leboeuf$^1$, A. G. Monastra$^2$, and A. Rela\~no$^3$}

\affiliation{
$^1$Laboratoire de Physique Th\'eorique et Mod\`eles Statistiques,
B\^at. 100, \\ Universit\'e de Paris-Sud, 91405 Orsay Cedex, France\\
$^2$Institut f\"ur Theoretische Physik, Technische
Universit\"at Dresden, 01062 Dresden, Germany\\
$^3$Departamento de F\'{\i}sica At\'omica, Molecular y Nuclear,
Universidad Complutense de Madrid, E-28040 Madrid, Spain}

%\date{\today, \RightText}
\date{\today}

\begin{abstract} 
We present a theory that accurately describes the counting of excited states
of a noninteracting fermionic gas. At high excitation energies the results
reproduce Bethe's theory. At low energies oscillatory corrections to the
many--body density of states, related to shell effects, are obtained. The
fluctuations depend non-trivially on energy and particle number. Universality
and connections with Poisson statistics and random matrix theory are
established for regular and chaotic single--particle motion.
\end{abstract}


\maketitle

The level density is a characteristic property of every many--body
quantum mechanical system. Its precise determination is often a key
ingredient in the calculation of different processes, like compound
nuclear decay rates, yields of evaporation residues to populate exotic
nuclei, or thermonuclear rates in astrophysical processes. The first
and main step towards the understanding of the density was given by
H. Bethe more than sixty years ago. He considered $A$ noninteracting
fermions moving in a mean--field potential. By this model he showed
that the number of excited states of the many--body (MB) system
contained in a small energy window $dQ$ at energy $Q$, $\rmb (A,Q)
dQ$, is \cite{bethe}
\begin{equation} \label{bethe}
\rmb (A,Q) = \frac{1}{\sqrt{48}\ Q} \exp \sqrt{\frac{2}{3}\pi^2 \rhob \ Q} \ .
\end{equation}
Here $\rhob$ is the average single--particle (SP) density of states at Fermi
energy $\ef$, and $Q$ is measured with respect to the ground state energy. For
a given potential, $\rhob$ is in general a function of $A$. This expression
assumes that the excitation energy is, on the one hand, large compared to the
SP mean level spacing $\delta = \rhob^{-1}$ and, on the other, small compared
to $\ef$ (degenerate gas approximation). Most of the present knowledge
concerning the density of states is based on refinements of Bethe's result.

In Eq.~(\ref{bethe}) all the information concerning the SP spectrum is encoded
in a unique parameter, $\rhob$, that describes its average behavior. The exact
dependence of $\rmb$ on $A$ and $Q$ is sensitive, however, to the detailed
arrangement of the SP energy levels around the Fermi energy. After Bethe's
work, more accurate calculations of $\rmb$ were made \cite{bloch}. Schematic
shell corrections related to a periodic fluctuation of the SP density were
computed in \cite{ros} (see also \cite{bm}), introducing the so--called
backshifted Bethe formula. Several modifications of Eq.(\ref{bethe}) have been
done to match the experimental results. These models take into account, for
example, shell effects, pairing corrections and residual interactions
\cite{para}, introducing a multitude of coexisting phenomenological
parameterizations. The present status of the understanding do not allow to
draw a clear theoretical picture of the functional dependence of $\rmb$ with
$A$ and $Q$.

Within an independent particle model we demonstrate, through a general theory,
the role played by shell effects in the MB density of states. Superimposed to
the smooth growth of the density described by Eq.(\ref{bethe}), there are
oscillatory corrections. To lowest order, these corrections are directly
related to energy fluctuations of the system, that may be expressed in terms
of sums over the classical periodic orbits. We make a statistical analysis of
the density fluctuations, in particular of their probability distribution,
typical size, and temperature dependence \cite{note}. We show that the
amplitude of the shell corrections grows linearly at low temperatures, and
decays as $T^{-1}$ when $T \gg E_c /2\pi^2$, where $E_c$ is the energy
conjugate to the time of flight across the system (temperature is measured in
units of Boltzmann constant). The statistical properties of the density
fluctuations strongly depend on the regular or chaotic nature of the
underlying classical SP motion. Universal properties for each class (regular
or chaotic) are found for $T \ll E_c /2\pi^2$. Besides the fluctuations, we
find and discuss corrections to the smooth part of the density in the exponent
of Eq.(\ref{bethe}). The present analysis considers general properties of
fermionic systems. Applications to the specific problem of the nuclear density
and to the interpretation of the experimental results will be done in a
separate contribution.

The MB density of states is defined as
\begin{equation} \label{rhodef}
\rmb (A,E) = \sum_N \sum_j \delta (A- N) \ \delta \left( E - E_{N,j} \right) \ ,
\end{equation}
where $E_{N,j}=\sum_i n_{j,i} \epsilon_i$ is the energy of the $j$-th SP
configuration of $N = \sum_i n_{j,i}$ particles, $\epsilon_i$ the SP energies,
and $n_{j,i}=0,1$ are the occupation numbers.

When the SP spectrum consists of equidistant levels separated by
$\delta$, e.g. the spectrum of a one dimensional harmonic oscillator,
an exact answer to the counting problem exists. The MB excitation
energies are integer multiples of $\delta$, each level having a
nontrivial degeneracy. It is easy to see that, for a given level
defined by an integer, the degeneracy is the same as the number of
ways into which the integer can be decomposed as a sum of integers
(the partition number). An asymptotic approximation to this
well--known mathematical problem was obtained by Hardy and Ramanujan,
and later on Rademacher found a convergent series
\cite{hrr}. Expressing it in terms of an expansion in terms of $\rhob
Q$, the result reads
\begin{eqnarray} \label{shr2}
&& \log \left( \rmb^{\text {\tiny HO}}/\rhob \right) = \sqrt{ \frac{2}{3}
\pi^2 \rhob \ Q } - \log \left( \sqrt{48} \ \rhob Q \right) - \\ && \ \
\frac{\pi^2 + 72}{24 \sqrt{6} \pi} (\rhob Q)^{-\frac{1}{2}} - \left(\frac{3}{4
\pi^2}- \frac{1}{24} \right) (\rhob Q)^{-1} + {\cal O} ((\rhob
Q)^{-\frac{3}{2}}) \nonumber \ .
\end{eqnarray}
The first two terms of this expansion reproduce Bethe's formula.

For an arbitrary SP spectrum the computation of the density of excited
states of a fermionic system is a difficult combinatorial problem for
which no exact solution is known. We expect, however, that
Eq. (\ref{shr2}) gives correctly the contribution of the average part
of the SP spectrum. What we are seeking here are the variations of the
MB density due to fluctuations of the SP spectrum with respect to a
perfectly ordered spectrum. A convenient way to express an approximate
solution is by means of an inverse Laplace transform of
Eq.(\ref{rhodef}). A saddle point approximation of the resulting
integrals yields \cite{bm,fow}
\begin{equation} \label{rhosp}
\rmb (A,E) = {\rm e}^{\sca (\mu,T)}/ \ 2\pi \sqrt{|\dc (\mu,T)|} \ ,
\end{equation}
valid in the degenerate gas approximation. The dependence on $A$ and $E$ in
Eq.~(\ref{rhosp}) arises from the saddle point conditions that fix the value
of the chemical potential $\mu$ and temperature $T$ of the gas
\begin{equation} \label{spc}
\nc (\mu, T) = A \ , \;\;\;\;\;\;\;\;\;\;\;\;\;\;\; \ec (\mu, T) = E \ . 
\end{equation}
The functions $\nc=-\partial \Omega/\partial \mu|_{\scriptscriptstyle T}$ and 
\begin{equation} \label{u}
\ec (\mu,T) = \Omega (\mu,T) + \mu \nc (\mu,T) + T \sca (\mu,T)
\end{equation}
are the particle number and energy functions of the gas, respectively,
with the entropy $\sca (\mu,T) =-\partial \Omega/\partial T|_\mu$ and
the grand potential $\Omega = - T \int d\epsilon \rho (\epsilon) \log
\left[1+{\rm e}^{(\mu-\epsilon)/T}\right]$, where $\rho (\epsilon) =
\sum_i \delta (\epsilon-\epsilon_i)$ is the SP density of states. In
terms of the previous functions, the determinant in Eq.(\ref{rhosp})
is defined as $\dc (\mu,T) = T^3 \left[ \partial \nc/\partial
\mu|_{\scriptscriptstyle T} \
\partial \ec/\partial T|_\mu - \partial \nc/\partial T|_\mu \ \partial
\ec/\partial \mu|_{\scriptscriptstyle T} \right]$. All the necessary quantities 
involved in the computation of $\rmb$ are therefore defined from
$\Omega (\mu,T)$.

The task consists in the evaluation of the entropy and the determinant at
values of $\mu$ and $T$ that satisfy the conditions (\ref{spc}). As $A$ and
$E$ are varied in Eqs.(\ref{spc}), in general $\mu$ and $T$ do not have a
smooth and gentle behavior, because the functions $\nc$ and $\ec$ may have
sudden changes due to the discrete nature of the SP spectrum. The difficulty
lies in a proper treatment of the smooth part of the variations as well as the
fluctuations with respect to it. To this purpose it is convenient to use the
semiclassical approximation \cite{gutz}. The SP density of states is
decomposed into the sum of a smooth term ($\rhob$, given by a Thomas Fermi
approximation or Weyl series) plus oscillatory terms $(\rhot)$, $\rho = \rhob
+ \rhot$. Taking into account only the smooth part of the SP density of
states, the calculation of $\rmb$ is relatively straightforward and leads to
Bethe's formula. The oscillatory part depends on the primitive classical
periodic orbits $p$ (and their repetitions $r$), $\rhot = 2 \sum_p
\sum_{r=1}^{\infty} A_{p,r} \cos \left[ r S_p/\hbar+ \nu_{p,r} \right]$. Each
orbit is characterized by its action $S_p$, stability amplitude $A_{p,r}$, and
Maslov index $\nu_{p,r}$. When inserted into the definition of the grand
potential, and after integration with respect to the energy, the fluctuating
part of $\Omega$ is given by \cite{ruj}
\begin{equation} \label{grandosc}
\widetilde{\Omega} (\mu,T)= 2 \hbar^2 \sum_p \sum_{r=1}^{\infty} \frac{A_{p,r}
~ \kappa ( \frac{r ~ \tau_p}{\taut})}{r^2 ~ \tau_p^2} \cos \left(
\frac{\scriptstyle r S_p}{\scriptstyle \hbar}+ \nu_{p,r} \right).
\end{equation}
Here $\tau_p$ is the period of the periodic orbit, and $\kappa (x) = x/\sinh
(x) $ is a temperature factor that introduces the time scale $\taut =
\hbar/(\pi T)$ conjugate to the temperature. This expression describes the
departures of $\Omega$ with respect to its mean behavior due to the
fluctuations of the SP spectrum. By simple derivation of the grand potential
the smooth and fluctuating part of all other thermodynamic functions can be
obtained. Intensive functions like the chemical potential and the temperature
should also be decomposed. We define their smooth parts by the implicit
equations
\begin{equation} \label{spc2}
\ncb (\mub, \tb) = A \ , \;\;\;\;\;\;\;\;\;\;\;\;\;\;\; \ecb (\mub,
\tb) = E \ ,
\end{equation}
whereas the fluctuating parts are $\mut = \mu - \mub$ and $\tti = T -
\tb$.

Using Eqs.(\ref{spc}) and the decomposition of the different
functions, Eq.(\ref{u}) may be rewritten as
\begin{equation} \label{s1}
(\tb + \tti) \sca = E - \ob (\mu,T) - \ot (\mu,T) - (\mub + \mut) A \ .
\end{equation}
If $A$ is large, the smooth part of the grand potential can be expanded in
power series of $\mut$ and $\tti$. Keeping terms up to first order, using that
in the degenerate gas approximation $\ob (\mub,\tb) \simeq \ob (\mub,0) -
(\pi^2/6) \rhob \ \tb^2$, and neglecting variations of $\rhob$, Eq.(\ref{s1})
becomes
\begin{eqnarray} \label{s2}
(\tb + \tti) \sca &=& E - \ob (\mub,0) + \frac{\pi^2}{6} \rhob \ \tb^2 +
\overline{\sca} (\mub,\tb) \ \tti - \nonumber \\ && \ot (\mu,T) - \mub A \ .
\end{eqnarray}
This equation is valid for an arbitrary temperature. At zero temperature, the
energy of the system is the ground state energy $E_0$, and Eq.(\ref{s2})
reduces to
\begin{equation} \label{s3}
0 = E_0 - \ob (\mub,0) - \ot (\mu_0,0) - \mub A \ .
\end{equation}
$\mub$ remains constant if variations of $\rhob$ are neglected, but the
fluctuating part still depends on temperature. As a consequence the chemical
potentials are slightly different. The idea is to subtract Eq.(\ref{s3}) from
(\ref{s2}). Before doing that we first need to properly analyze some of the
terms. The difference $E-E_0=Q$ is the excitation energy of the gas, as well
as the term $(\pi^2/6) \rhob \ \tb^2 =Q$. The latter equation follows from the
dependence of the energy on temperature in the degenerate gas approximation,
$\ecb (\mub,\tb) \simeq \ecb (\mub,0) + (\pi^2/6) \rhob \ \tb^2$, and the
stationary phase condition (\ref{spc2}) for $\ecb$ (and the corresponding one
at zero temperature). It allows to express the temperature in terms of $Q$.
Similarly, $\mub$ is obtained by inversion of $\ncb (\mub,\tb)$ in
(\ref{spc2}) which, in the degenerate gas approximation and neglecting
variations of $\rhob$, is independent of $\tb$.

Returning to Eq.(\ref{s2}), the term $\tti (\sca - \overline{\sca} (\mub,\tb))
= \tti \widetilde{\sca}$ is of second order, and is therefore neglected.
After subtraction of Eq.(\ref{s3}) and expressing $\tb$ in terms of $Q$, the
entropy may be expressed as
\begin{eqnarray} \label{sdef}
\sca (A,Q) &=& \sqrt{\frac{2}{3} \pi^2 \rhob \ Q} + \frac{1}{\tb}
\left[ \ot (\mub,0) - \ot (\mub,\tb) \right] \\
&\approx& \sqrt{\frac{2}{3} \pi^2 \rhob \ \left[ Q + \ot (\mub,0) -
\ot (\mub,\tb) \right]} \ , \label{sdef2}
\end{eqnarray}
where, to lowest order, $\ot$ has been evaluated at $(\mub, \tb)$ (a similar
though improved accuracy is obtained using $\mu$ and $T$, cf. Eq.(\ref{hro})
below). Equation (\ref{sdef2}) is valid if $\ot (\mub,0) - \ot (\mub,\tb) \ll
Q$, and puts the result under the form of a backshifted Bethe formula.

A more physical interpretation of Eqs.(\ref{sdef}) and (\ref{sdef2}), that
expresses the result in an entirely microcanonical language, is provided by
the following connection. It has been shown previously \cite{lm3,sev} that, to
leading order in an expansion in terms of $\mut$, $\ot (\mub,\tb) = \ect
(A,Q)$, where $\ect (A,Q)$ are the (shell) fluctuations of the energy of the
gas at a fixed number of particles and excitation energy. The shell
corrections of Eq.(\ref{sdef2}) are therefore directly related to the
fluctuations of the energy of the system.

A similar analysis can be performed for the determinant $\dc (\mu,T)$ in
Eq.(\ref{rhosp}). To leading order we find
\begin{equation} \label{det}
\dc = \frac{\pi^2}{3} \rhob^2 \ \tb^4 = \frac{12}{\pi^2} \ Q^2 \ ,
\end{equation}
recovering the same form obtained by Bethe ($2\pi \sqrt{\dc}=\sqrt{48}
\ Q$, see Eq.(\ref{bethe})). There exist oscillatory corrections to
$\dc$, but these are exponentially small compared to those associated
with Eq.(\ref{sdef}), and we neglect them.

Equation (\ref{sdef}) is the central result of the paper. It expresses,
together with Eq.(\ref{det}), the MB density of states in terms of the
particle number $A$ and excitation energy $Q$. It decomposes $\sca$ into a
smooth and a fluctuating part, $\sca = \scab + \scat$. The smooth part is
given by the usual Bethe's result (corrections to it may be obtained by
keeping higher order terms in $\mut$ and $\tti$ in the expansion). The novelty
in Eq.(\ref{sdef}) is the additional contribution of oscillatory corrections,
$\scat = [\ot (\mub,0) - \ot (\mub,\tb)]/\tb$. They describe shell corrections
to the MB density. $\widetilde{\Omega}$ depends on temperature only through
the function $\kappa$. The effect of this function is to exponentially
suppress the contribution of periodic orbits whose period $\tau_p \gg \taut$
\cite{ruj,lm3}. Since there is no suppression at $T=0$ (because $\taut
\rightarrow \infty$), only orbits whose period $\tau_p \gtrsim \taut$
contribute to the difference $\ot (\mub,0) - \ot (\mub,\tb)$. At temperatures
such that $\taut \ll \tmin$, where $\tmin$ is the period of the shortest
periodic orbit of the system, the term $\ot (\mub,\tb)$ becomes exponentially
small, and only $\ot (\mub,0)$ remains. The shell correction $\scat$ therefore
decays as $\tb^{-1}$. This decay contrasts with the more common exponential
damping observed in other thermodynamic quantities \cite{lm3}.

The behavior of $\widetilde{\Omega}$ (and therefore of $\scat$) strongly
depends on whether $\mu$ or $T$ are varied. As discussed above, a temperature
variation modifies the prefactors of the summands in $\widetilde{\Omega}$
(through the function $\kappa$), and therefore produces gentle variations of
$\scat$. In contrast, $A_{p,r}$, $\tau_p$ and $S_p$ depend on $\mu$. For large
values of $\mu$, $S_p \gg \hbar$ and the dominant variation with the particle
number (or any other parameter that modifies the actions) comes from the
argument of the cosine function in $\widetilde{\Omega}$. Rapid oscillations of
$\scat$ are therefore generically expected when the number of particles is
varied.

A clearer picture of how the fluctuations behave may be obtained through a
statistical analysis. There are two relevant SP scales in the analysis, the
mean spacing $\delta$ and the energy conjugate to the shortest periodic orbit,
$E_c = h/\tmin$. The ratio $g = E_c/\delta$ is typically much larger than $1$
(for instance, $g\sim A^{2/3}$ in a three dimensional cavity). The most simple
property of the fluctuations is $\langle \scat \rangle =0$, where the brackets
denote an average over a suitable chemical potential window. This result is
valid only to first order in the expansion; it can be shown that higher order
terms contribute to a non--zero average. The next non--trivial statistical
property is the variance $\langle \scat^2 \rangle$, that may be computed using
Eq.(\ref{grandosc}). The result is $\langle \scat^2 \rangle = (1/2)
\int_{0}^\infty d x \ K(x,\xh) \left[ 1 - \kappa (x) \right]^2 /x^4$, where
$K(x,\xh)$ is the rescaled form factor of the SP spectrum (cf Eq.(36) in
Ref.\cite{lm3}). The latter function depends on the rescaled Heisenberg time
$\xh=h \rhob/\taut$. It describes system--dependent features for $x$ of the
order of $\xmin = \tmin/\taut$, while it is believed to be universal for $x
\gg \xmin$. The universality class depends on the regular or chaotic nature of
the dynamics, and on its symmetry properties. Taking into account the basic
properties of $K(x,\xh)$, in chaotic systems we find three different regimes
for $\langle \scat^2 \rangle$ as a function of $\tb$:

\noindent (i) Low temperatures $2\pi^2 \tb \ll \delta$. In this regime
$\langle \scat^2 \rangle = c_4 \pi^2 \tb/\delta$, where $c_4 = 0.0609...$.

\noindent (ii) Intermediate temperatures $\delta \ll 2\pi^2 \tb \ll E_c$. In
this regime the size of the fluctuations saturate at a universal
constant $\langle \scat^2 \rangle = c_3/\beta$, where $c_3 =
0.1023\ldots$ and $\beta = 1 (2)$ for systems with (without)
time--reversal invariance.

\noindent (iii) High temperatures $2\pi^2 \tb \gg E_c$. The size of the
fluctuations decreases with excitation energy, $\langle \scat^2 \rangle =
\langle \ot^2 (\mub,0) \rangle [ 1 - 8 \ {\rm e}^{-2 \pi^2 \tb/E_c} ]/ \tb^2$.
After an exponential transient, a power--law decay $\langle \scat^2
\rangle^{1/2} \propto \tb^{-1}$ is obtained.

The situation is different in integrable systems, where only two regimes are
found. At low temperatures the result is identical to that of chaotic systems.
The difference is that now the growth extends up to much higher temperatures,
$2 \pi^2 \tb \approx E_c$, without saturation. At that temperature the
variance of the fluctuations is of order $g$. In integrable systems, the
maximum amplitude of the fluctuations is therefore reached at $2 \pi^2 \tb
\approx E_c$, and its typical size is much larger than in chaotic systems. At
high temperatures $2 \pi^2 \tb \gg E_c$ the decay is almost identical to that
of chaotic systems (the coefficient 8 is replaced by a 12).

As shown elsewhere \cite{lm3}, $\ot (\mub,\tb)$ is dominated, at any $\tb$, by
the shortest classical periodic orbits. In contrast, the difference $\ot
(\mub,0)-\ot (\mub,\tb)$ depends on orbits whose period $\tau_p \geqa \taut$.
For temperatures $2 \pi^2 \tb \ll E_c$ the statistical properties of these
orbits are universal, and correspondingly the probability distribution
function of $\scat$ is expected to be universal, in the sense that at a given
temperature it should only depend on the nature of the underlying classical
dynamics (regular or chaotic), and the symmetries of the system. This
statement is supported by the fact that in the limit $\tb \rightarrow 0$,
$\scat \approx - \partial \ot (\mub,\tb) /\partial \tb$. The probability
distribution of the latter quantity was studied in Ref.\cite{lm3}; it was
shown that it coincides at low temperatures with that obtained from a Poisson
spectrum for integrable systems and from a random matrix spectrum for chaotic
ones. As the temperature is raised, the universality of the probability
distribution of $\scat$ will be lost for temperatures of the order or greater
than $E_c$, where system specific features are revealed.

\begin{figure}
\includegraphics[width=8.5cm,height=7.0cm]{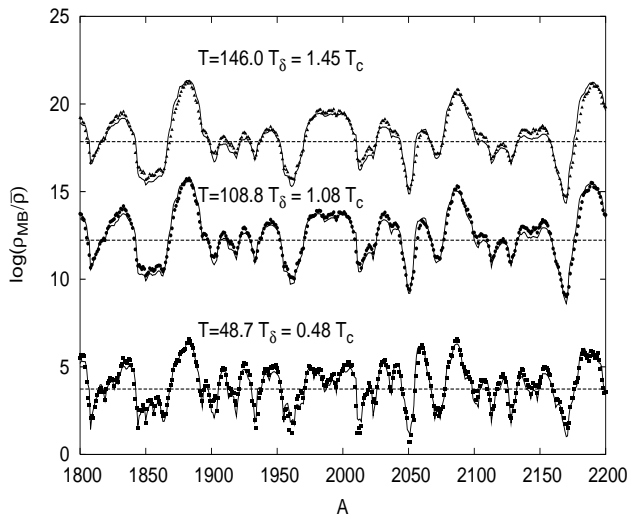}
\caption{The logarithm of the many-body density of states of $A$
     noninteracting fermions in a rectangular billiard of sides $a = \sqrt{(1
     + \sqrt{5})/2}$ and $b=a^{-1}$. Dots: numerical computation at three
     different excitation energies; solid curves:  theoretical prediction
     (\ref{hro}); dashed curves: smooth part Eq.(\ref{shr2}).}
\end{figure}

We have checked some of our predictions by a direct numerical counting of the
MB density of states in a particular system. Figure 1 shows the results
obtained for a gas of about 2000 fermions contained in a two--dimensional
rectangular cavity, an integrable system. For each number of particles we
compute the MB density of states at three different temperatures, measured in
units of $T_{\delta}=\delta/2 \pi^2$ and $T_c =E_c /2 \pi^2$. The theoretical
curve is computed according to the expression
\begin{equation} \label{hro}
\log ( \rmb/\rhob ) = \log ( \rmb^{\text{\tiny HO}}/\rhob ) + [ \ot
(\mu,0) - \ot (\mu,\tb) ]/\tb \ ,
\end{equation}
where $\ot$ is given by the periodic orbits of the rectangle. Though the
results are similar, $\mu$ instead of $\mub$ is used to obtained a more
accurate description of the numerical data. A systematic deviation is observed
between theory and numerics if the corrections arising from the exact
expression Eq.(\ref{shr2}) are not included. Notice the extremely good
accuracy of this equation, either for the average value of the density as well
as for the fluctuations.

\noindent The Laboratoire de Physique Th\'eorique et Mod\`eles Statistiques is
an Unit\'e de recherche de l'Universit\'e Paris XI associ\'ee au CNRS.


\begin{thebibliography}{10}
\bibitem{note} All over the text, we may either refer to the excitation energy
$Q$ of the system or to its temperature $T$.
\end{thebibliography}
\end{document}